\pdfoutput=1
\documentclass[12pt,a4paper]{article}

\usepackage{ifthen}
\newboolean{pdflatex}
\setboolean{pdflatex}{true}

\newboolean{articletitles}
\setboolean{articletitles}{true}

\newboolean{uprightparticles}
\setboolean{uprightparticles}{false}

\def\paperauthors{LHCb collaboration} 
\def\paperasciititle{Nuclear modification factor of neutral pions in the forward and backward regions in p-Pb collisions}
\def\papertitle{Nuclear modification factor of neutral pions in the forward and backward regions in $p$-Pb collisions}

\def\paperkeywords{{High Energy Physics}, {LHCb}}
\def\papercopyright{\the\year\ CERN for the benefit of the LHCb collaboration}
\def\paperlicence{CC BY 4.0 licence}
\def\paperlicenceurl{https://creativecommons.org/licenses/by/4.0/}

\usepackage[top=1in, bottom=1.25in, left=1in, right=1in]{geometry}

\usepackage{microtype}
\usepackage{lineno}
\usepackage{xspace}
\usepackage{caption}

\usepackage{graphicx}  
\usepackage{color}
\usepackage{colortbl}

\usepackage{amsmath} 
\usepackage{amssymb}
\usepackage{amsfonts}
\usepackage{upgreek} 

\newcommand*\patchAmsMathEnvironmentForLineno[1]{%
\expandafter\let\csname old#1\expandafter\endcsname\csname #1\endcsname
\expandafter\let\csname oldend#1\expandafter\endcsname\csname
end#1\endcsname
 \renewenvironment{#1}%
   {\linenomath\csname old#1\endcsname}%
   {\csname oldend#1\endcsname\endlinenomath}%
}
\newcommand*\patchBothAmsMathEnvironmentsForLineno[1]{%
  \patchAmsMathEnvironmentForLineno{#1}%
  \patchAmsMathEnvironmentForLineno{#1*}%
}
\AtBeginDocument{%
\patchBothAmsMathEnvironmentsForLineno{equation}%
\patchBothAmsMathEnvironmentsForLineno{align}%
\patchBothAmsMathEnvironmentsForLineno{flalign}%
\patchBothAmsMathEnvironmentsForLineno{alignat}%
\patchBothAmsMathEnvironmentsForLineno{gather}%
\patchBothAmsMathEnvironmentsForLineno{multline}%
\patchBothAmsMathEnvironmentsForLineno{eqnarray}%
}

\usepackage{hyperxmp}

\usepackage[pdftex,
            pdfauthor={\paperauthors},
            pdftitle={\paperasciititle},
            pdfkeywords={\paperkeywords},
            pdfcopyright={Copyright (C) \papercopyright},
            pdflicenseurl={\paperlicenceurl}]{hyperref}

\usepackage[bottom,flushmargin,hang,multiple]{footmisc}

\usepackage[all]{hypcap}

\usepackage{xspace} 
\usepackage{upgreek}

\def\lhcb   {\mbox{LHCb}\xspace}

\def\Ppi         {\ensuremath{\pi}\xspace}
\def\pion   {{\ensuremath{\Ppi}}\xspace}
\def\piz    {{\ensuremath{\pion^0}}\xspace}
\def\PD      {\ensuremath{D}\xspace}                 
\def\D       {{\ensuremath{\PD}}\xspace}
\def\Dz      {{\ensuremath{\D^0}}\xspace}
\def\Pp      {\ensuremath{p}\xspace}
\def\proton      {{\ensuremath{\Pp}}\xspace}

\newcommand{\nospaceunit}[1]{\ensuremath{\text{#1}}}       
\newcommand{\aunit}[1]{\ensuremath{\text{\,#1}}}       

\newcommand{\tev}{\aunit{Te\kern -0.1em V}\xspace}
\newcommand{\gev}{\aunit{Ge\kern -0.1em V}\xspace}
\newcommand{\mev}{\aunit{Me\kern -0.1em V}\xspace}

\def\mub{\ensuremath{\,\upmu\nospaceunit{b}}\xspace}

\def\sqsnn {\ensuremath{\protect\sqrt{s_{\scriptscriptstyle\text{NN}}}}\xspace}
\def\pt         {\ensuremath{p_{\mathrm{T}}}\xspace}

\newcommand{\lhcborcid}[1]{\href{https://orcid.org/#1}{\hspace*{0.1em}\raisebox{-0.45ex}{\includegraphics[width=1em]{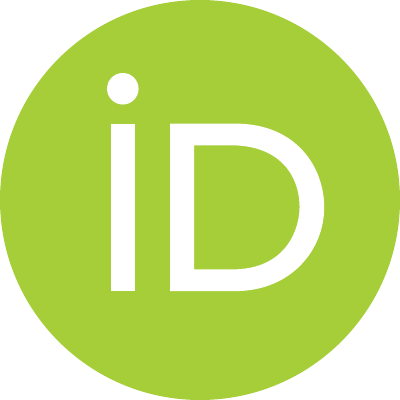}}}}

\usepackage{cite}
\usepackage{mciteplus}

\begin{document}

\renewcommand{\thefootnote}{\fnsymbol{footnote}}
\setcounter{footnote}{1}

\begin{titlepage}
\pagenumbering{roman}

\vspace*{-1.5cm}
\centerline{\large EUROPEAN ORGANIZATION FOR NUCLEAR RESEARCH (CERN)}
\vspace*{1.5cm}
\noindent
\begin{tabular*}{\linewidth}{lc@{\extracolsep{\fill}}r@{\extracolsep{0pt}}}
\ifthenelse{\boolean{pdflatex}}
{\vspace*{-1.5cm}\mbox{\!\!\!\includegraphics[width=.14\textwidth]{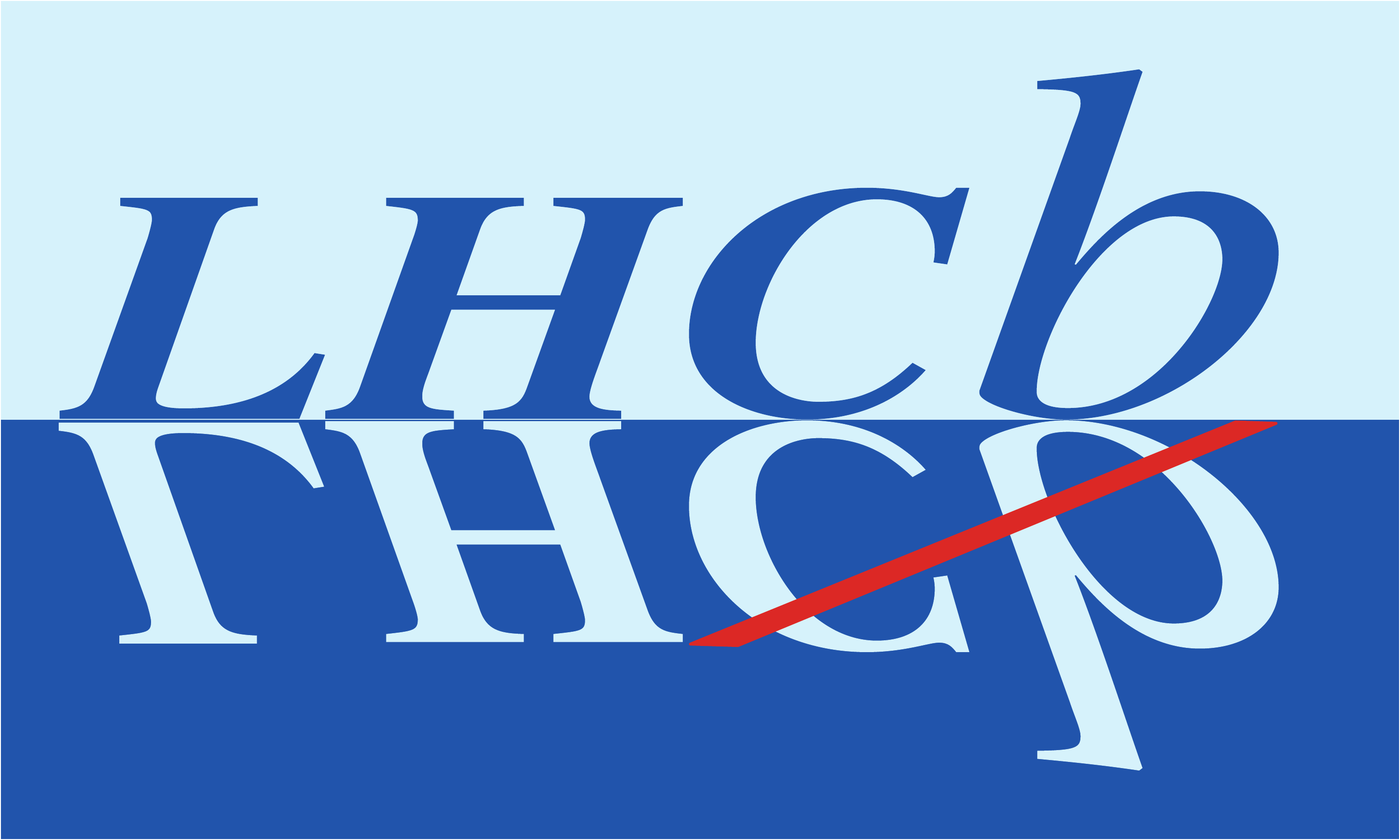}} & &}%
{\vspace*{-1.2cm}\mbox{\!\!\!\includegraphics[width=.12\textwidth]{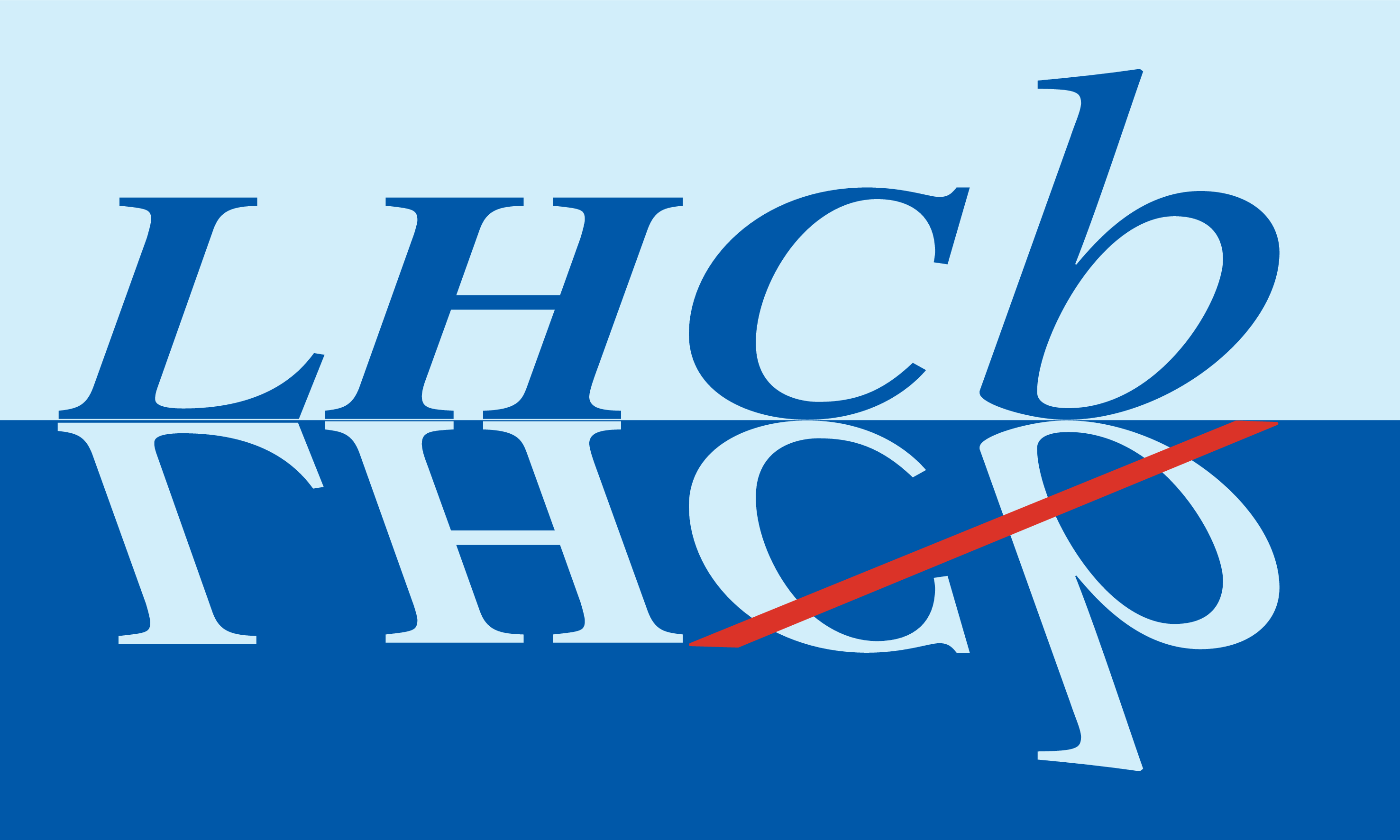}} & &}%

\\
 & & CERN-EP-2022-050 \\  
 & & LHCb-PAPER-2021-053 \\  
 & & 24 July 2023 \\ 
 & & \\
\end{tabular*}

\vspace*{4.0cm}

{\normalfont\bfseries\boldmath\huge
\begin{center}
  \papertitle 
\end{center}
}

\vspace*{2.0cm}

\begin{center}
\paperauthors\footnote{Authors are listed at the end of this Letter.}
\end{center}

\vspace{\fill}

\begin{abstract}
  \noindent
The nuclear modification factor of neutral pions is measured in proton-lead
collisions collected at a center-of-mass energy per nucleon of
\mbox{$8.16\,\tev$} with the LHCb detector. The \piz production cross section is
measured differentially in transverse momentum ($\pt$) for
\mbox{$1.5<\pt<10.0\,\gev$} and in center-of-mass pseudorapidity ($\eta_{\rm
c.m.}$) regions \mbox{$2.5<\eta_{\rm c.m.}<3.5$} (forward) and
\mbox{$-4.0<\eta_{\rm c.m.}<-3.0$} (backward) defined relative to the proton
beam direction.  The forward measurement shows a sizable suppression of $\piz$
production, while the backward measurement shows the first evidence of $\piz$
enhancement in proton-lead collisions at the LHC. Together, these measurements
provide precise constraints on models of nuclear structure and particle
production in high-energy nuclear collisions.
  
\end{abstract}

\vspace*{2.0cm}

\begin{center}
  Published in Physical Review Letters {\bf 131} (2023) 042302
\end{center}

\vspace{\fill}

{\footnotesize 

\centerline{\copyright~\papercopyright. \href{\paperlicenceurl}{\paperlicence}.}}
\vspace*{2mm}

\end{titlepage}


\newpage
\setcounter{page}{2}
\mbox{~}

\renewcommand{\thefootnote}{\arabic{footnote}}
\setcounter{footnote}{0}

\cleardoublepage


\pagestyle{plain}
\setcounter{page}{1}
\pagenumbering{arabic}

Neutral pion production is an important probe of nuclear effects in heavy ion
collisions. In proton-lead ($p$-Pb) collisions at the Large Hadron Collider
(LHC), $\piz$ production is particularly sensitive to cold nuclear matter (CNM)
effects on the initial state of the bound nucleons in the colliding nucleus.
Pion production in $p$-Pb collisions can be described using the collinear
factorization framework where CNM effects are encoded into nuclear parton
distribution functions
(NPDFs)~\cite{Kovarik:2019xvh,Eskola:2016oht,Kovarik:2015cma,deFlorian:2011fp}
and the parton-parton collision dynamics are described by perturbative quantum
chromodynamics (PQCD)~\cite{Collins:1989gx}. These NPDFs are determined using
fits to data and are poorly constrained for partons with momentum fraction $x$
smaller than about $10^{-4}$. At low $x$ and low momentum transfer $Q$, the
parton number density may become so large that parton saturation effects become
sizable. Parton saturation would lead to nonlinear evolution of parton densities
and is often described using the color glass condensate (CGC) effective field
theory~\cite{Kowalski:2007rw}. Measurements of $\piz$ production in $p$-Pb
collisions at forward and backward rapidities with the LHCb detector can provide
constraints on NPDFs for $x$ between $10^{-6}$ and $10^{-1}$, potentially
helping identify the onset of parton saturation effects~\cite{Morreale:2021pnn}.

Measurements of angular correlations in small-collision systems at the
LHC~\cite{CMS:2010ifv,ALICE:2012eyl,ATLAS:2012cix,CMS:2012qk,LHCb-PAPER-2015-040}
and the Relativistic Heavy Ion Collider
(RHIC)~\cite{PHENIX:2013ktj,PHENIX:2018lia} show evidence of
collective flow, which cannot be described using only collinear
factorization and NPDFs. In addition, the LHCb experiment recently
measured the nuclear modification factor of inclusive charged
particles in $p$-Pb collisions at a center-of-mass energy per
nucleon of \mbox{$\sqrt{s_{\rm NN}}=5\,\tev$}, observing a much larger
enhancement at backward pseudorapidities than predicted by NPDF
calculations~\cite{LHCb-PAPER-2021-015}. Similar enhancements have
been observed in charged-particle production at
RHIC~\cite{PHENIX:2003qdw,STAR:2003pjh,BRAHMS:2003sns,PHENIX:2019gix}
and have been attributed to the Cronin
effect~\cite{Cronin:1974zm}. The Cronin effect is often described as
initial-state multiple scattering within the
nucleus~\cite{Accardi:2002ik}, although radial collective flow and
final-state recombination could produce similar
enhancements~\cite{Lee:1990sk,Hwa:2004zd}. However, measurements of
inclusive charged-particle and $\piz$ production at central rapidity
at the LHC show no evidence of a Cronin-like
enhancement~\cite{ALICE:2018vuu,CMS:2016xef,ALICE:2018vhm,ALICE:2021est},
potentially pointing to interplay between enhancing effects and
low-$x$ effects such as parton
saturation~\cite{Rezaeian:2012ye,Kharzeev:2003wz}.

Untangling the causes of the effects observed at RHIC and the LHCb
experiment requires measurements of the nuclear modification factor of
identified particles, such as neutral pions, over a broad rapidity
range. Multiple scattering in the nucleus would result in similar
enhancements for all particle species, while radial flow is expected
to have a larger transverse momentum ($\pt$) hardening effect on
higher mass particles included in charged-particle
measurements~\cite{Ayala:2006bc}. Alternatively, recombination is
expected to produce a particularly large baryon
enhancement~\cite{Hwa:2004zd}. The PHENIX collaboration recently
reported enhancements of $\piz$ production in small-collision systems,
observing a system size dependence qualitatively consistent with
radial flow, although still quantitatively consistent with NPDF
predictions~\cite{Acharya:2021tnd}. A study of $\piz$ production with
the LHCb detector will help differentiate between contributions from
nuclear parton density effects, initial state multiple scattering, and
final-state collective flow.

This Letter presents a measurement of the nuclear modification factor
of $\piz$ meson production in $p$-Pb collisions
at \mbox{$\sqrt{s_{\rm NN}}=8.16\,\tev$} with the LHCb detector. The
nuclear modification factor is given by
\begin{equation}
    R_{p\mbox{-}{\rm Pb}} \equiv \frac{1}{A}\frac{d\sigma_{p\mbox{-}{\rm Pb}}/d\pt}{d\sigma_{pp}/d\pt},
\end{equation}
where $A=208$ is the atomic mass number of the lead nucleus and
$d\sigma_{p\mbox{-}{\rm Pb}}/d\pt$ and $d\sigma_{pp}/d\pt$ are the $\piz$
differential production cross sections in $p$-Pb and proton-proton ($pp$)
collisions, respectively. The $\piz$ cross section is measured in $p$-Pb
collisions at \mbox{$\sqrt{s_{\rm NN}}=8.16\,\tev$} at both forward and backward
rapidities relative to the proton beam. The LHCb detector collected only a small
sample of unbiased $pp$ collisions at \mbox{$\sqrt{s}=8\,\tev$}, hence a $pp$
reference cross section is constructed by interpolating between measurements
performed using $pp$ collisions at \mbox{$\sqrt{s}=5$} and $13\,\tev$. The
nuclear modification factor is measured for \mbox{$1.5<\pt<10.0\,\gev$} in the
pseudorapidity regions \mbox{$2.5<\eta_{\rm c.m.}<3.5$} and
\mbox{$-4.0<\eta_{\rm c.m.}<-3.0$}, where $\eta_{\rm c.m.}$ is the
pseudorapidity in the center-of-mass frame related to the laboratory frame
pseudorapidity $\eta_{\rm lab}$ by \mbox{$\eta_{\rm c.m.}=\eta_{\rm lab}-0.465$}
in $p$-Pb collisions and \mbox{$\eta_{\rm c.m.}=\eta_{\rm lab}$} in $pp$
collisions (natural units are used throughout this Letter).

The LHCb detector is a single-arm forward spectrometer covering the
pseudorapidity range $2<\eta_{\rm lab}<5$, described in detail in
Refs. \cite{LHCb-DP-2008-001,LHCb-DP-2014-002}.  The detector includes a
high-precision tracking system consisting of a silicon-strip vertex detector
(VELO) surrounding the interaction region, a large-area silicon-strip detector
located upstream of a dipole magnet, and three stations of silicon-strip
detectors and straw drift tubes placed downstream of the magnet. Photons and
electrons are identified by a calorimeter system consisting of scintillating-pad
and preshower detectors, an electromagnetic (ECAL) and a hadronic calorimeter.
Particularly important for this analysis is the ECAL, which consists of
alternating layers of lead and scintillator and has an energy ($E$) resolution
of \mbox{$13.5\%/\sqrt{E/{\rm
GeV}}\oplus5.2\%\oplus(0.32\,\gev)/E$}~\cite{LHCb-DP-2020-001}. Simulated data
samples are used to model the detector response to neutral pion reconstruction.
In the simulation, $p$-Pb collisions are generated using
EPOS-LHC~\cite{Pierog:2013ria}, while $pp$ collisions are generated using
\textsc{Pythia}~\cite{Sjostrand:2007gs} with a specific LHCb
configuration~\cite{LHCb-PROC-2010-056}. Decays of unstable particles are
described by \textsc{EvtGen}~\cite{Lange:2001uf} and final-state radiation is
generated using \textsc{Photos}~\cite{davidson2015photos}. The interaction of
the generated particles with the detector, and its response, are implemented
using the \textsc{Geant4} toolkit~\cite{Allison:2006ve,Agostinelli:2002hh} as
described in Ref.~\cite{LHCb-PROC-2011-006}.

Proton-lead collisions are recorded in the forward configuration in which the
proton beam travels from the interaction region toward the spectrometer, and in
the backward configuration in which the lead ion beam travels toward the
spectrometer. The forward (backward) datasets correspond to an integrated
luminosity of \mbox{$328\pm9\,\mub^{-1}$} \mbox{($267\pm7\,\mub^{-1}$)}. The
forward configuration data are used to perform measurements for positive
$\eta_{\rm c.m.}$, while the backward sample is used for negative $\eta_{\rm
c.m.}$. All $p$-Pb events must have at least one reconstructed track in the
VELO, while the $pp$ samples consist of unbiased events.

Neutral pions are reconstructed using their decays to pairs of
photons. To ensure that both photons are in the detector acceptance,
$\piz$ candidates must have $2.5<\eta_{\rm lab}<4.0$. Photons may
reach the ECAL, or they may convert to electron-positron pairs in the
detector material. Photons that reach the ECAL or that convert in the
material downstream of the dipole magnet are reconstructed as energy
deposits (clusters) in the ECAL and are referred to as ECAL
photons. Photons that convert upstream of the dipole magnet are
referred to as converted photons. Only $\piz$ candidates reconstructed
from combinations of one ECAL photon and one converted photon are
considered. This reconstruction method provides better momentum
resolution than ECAL photon pairs alone and does not suffer from
systematic effects due to the small opening angle between $\piz$ decay
photons. ECAL photon clusters must contain at least two ECAL cells and
must be far from the extrapolated trajectories of all reconstructed
tracks. Converted photons are reconstructed from pairs of good-quality
tracks with $p>2\,\gev$.  To minimize the effects of energy loss via
bremsstrahlung radiation, only photons that convert downstream of the
VELO are considered.  Converted photons are required to have a small
invariant mass, with a maximum that varies based on the position of
the conversion vertex along the beam axis. The reconstructed converted
photon momentum is also required to point to a reconstructed primary
vertex. The ECAL photon must have $\pt>400\mev$ and the converted
photon must have $\pt>500\mev$.

\begin{figure}[tb]
\centering
\includegraphics[width=0.6\textwidth]{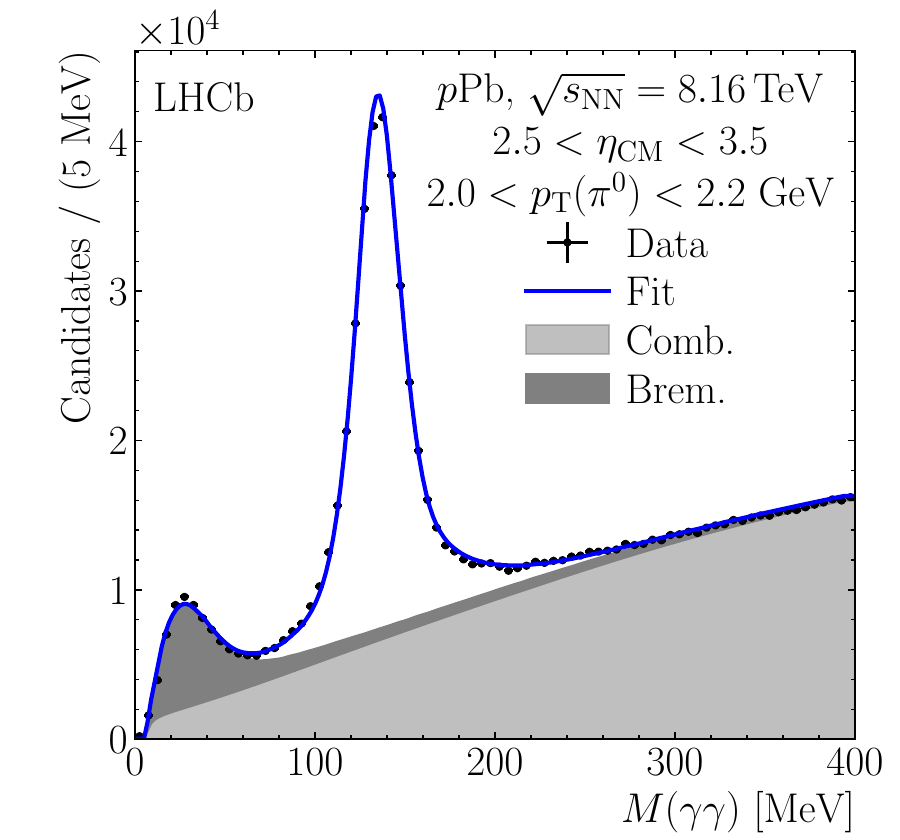}
\caption{Example $M(\gamma\gamma)$ distribution in forward $p$-Pb collisions at
\mbox{$\sqrt{s_{\rm NN}}=8.16\,\tev$} for \mbox{$2.5<\eta_{\rm c.m.}<3.5$} and
\mbox{$2.0<\pt(\piz)<2.2\,\gev$}. Fit results are overlaid with the total fit
shown as a solid line. The combinatorial (Comb.) and bremsstrahlung (Brem.)
backgrounds are shown as light and dark shaded regions, respectively.}
\label{fig:fitexample}
\end{figure}

An example $\piz$ candidate invariant mass [$M(\gamma\gamma)$]
distribution in one interval of $\pt$ and $\eta_{\rm c.m.}$ is shown in
Fig.\,\ref{fig:fitexample}. The $\piz$ yield is determined using a
binned maximum likelihood fit to the $M(\gamma\gamma)$
distribution. The $\piz$ signal is modeled by a two-sided Crystal
Ball function~\cite{Skwarnicki:1986xj}. The parameters describing the
tails of the signal distribution are determined from simulation, while
the Gaussian mean and width are left to vary in the fit to the
data. The combinatorial background is modeled using charged particles
as proxies for neutral pions. Charged particles reconstructed in the
tracking system are given the $\piz$ mass, and their decays to two
photons are simulated. The simulated photons are combined with
reconstructed ECAL photons to form background candidates. The mass
distribution of the proxy background candidates accurately describes
combinatorial backgrounds in simulation and is used as a background
model in the fit. An additional background component arises when a
converted photon is combined with its own bremsstrahlung radiation to
form a $\pi^0$ candidate, producing a peak at low mass. This
background is modeled by convolving the reconstructed converted
photon mass distribution with the sum of two positive half-normal
distributions. The width of the narrower half-normal distribution is
left to vary in the fit, while the larger width is fixed using
simulation.

The $\piz$ yields are corrected for the effects of the detector
response using simulation and an iterative unfolding procedure. First,
correction factors are calculated for each $\pt$ interval from
simulation and are used to correct the measured $\piz$ yields. The
corrected $\piz$ spectrum in data and the true $\piz$ spectrum in
simulation are then fit to Hagedorn
functions~\cite{PHENIX:2009gyd}. The ratio of these fits is used to
weight the true $\piz$ $\pt$ spectrum in simulation. The procedure is
then repeated with the weighted simulated data sample. The procedure
consistently converges after three iterations.

Differences in reconstruction efficiency between data and simulation
are measured and corrected for~\cite{LHCb-DP-2019-003}. The ECAL
photon reconstruction efficiency is the product of the cluster
reconstruction efficiency and the photon cluster identification
efficiency. The former is estimated with a tag-and-probe method using
converted photons. The tag electron must be matched to an ECAL cluster
and be identified as an electron. The cluster efficiency is then the
fraction of probe electron tracks with a matching ECAL cluster. The
photon cluster identification efficiency is evaluated by repeating the
photon reconstruction without any photon cluster identification
requirements in a subset of events in each data sample and measuring
the fraction of $\piz$ mesons that pass the requirements. The
converted-photon efficiency is measured using the ratio of the yields
of $\piz$ mesons reconstructed as an ECAL photon and a converted
photon to those reconstructed as two ECAL photons. This ratio is
measured for $1.0<\pt<1.5\,\gev$ where ECAL photons from $\piz$ decays
are always reconstructed as two separate clusters. The result is then
combined with the measured ECAL photon efficiency correction factors
to calculate a converted-photon efficiency correction. The corrected
efficiencies range from about $0.5\%$ at $1.5\,\gev$ to about $5\%$ at
$10\,\gev$.

The $\piz$ cross section in $pp$ collisions at $\sqrt{s}=8.16\,\tev$ is
estimated by interpolating between the measured $pp$ cross sections at
\mbox{$\sqrt{s}=5$} and \mbox{$\sqrt{s}=13\,\tev$}. The interpolation is
performed independently in each $\pt$ interval. The cross section is
interpolated using the functional form \mbox{$\sigma(s)=as^b$}. A linear
interpolation in $\sqrt{s}$ is also considered, as is a relative placement
interpolation method~\cite{CMS:2015ved}. In the latter, placement factors are
calculated using simulation assuming a linear or power-law dependence of the
cross section on $\sqrt{s}$. The placement factors are then applied to data to
determine the interpolated cross section.

The systematic uncertainties are summarized in Table~\ref{tab:systematics}. The
largest sources of systematic uncertainty come from the interpolation between
$pp$ cross sections and the $\piz$ fit model. The maximum variation of
interpolation results using the different $pp$ interpolation methods is taken as
the interpolation uncertainty. The fit model uncertainty is estimated by varying
the default fit parameters by their uncertainties from the simulation. The
variance of the resulting yields is taken as a systematic uncertainty. The fit
model uncertainty also includes a contribution from the background model, which
is estimated by repeating the fit using a polynomial background and taking the
difference relative to the default result as a systematic uncertainty. An
additional contribution to the fit model uncertainty is assigned to the yield
extraction method itself by using an alternative method, where a background-only
fit is performed in the mass region \mbox{$[0,50]\cup[250,400]\,\mev$}. The
background is then subtracted, and the background-subtracted distribution is
integrated over $[50,250]\,\mev$ to estimate the $\piz$ yield. The difference
between the alternative and default yield extraction methods is taken as a
systematic uncertainty.

Smaller systematic uncertainties come from unfolding, the luminosity
estimate, and the efficiency correction factors. The unfolding
uncertainty is estimated by using a closure test in simulation. The
unfolded yields agree with the true yields in simulation to within
about $1\%$ in most $\pt$ intervals. An additional unfolding
uncertainty arises from differences in $\piz$ $\pt$ resolution in data
and simulation. This difference is estimated to be less than $10\%$ by
comparing the fitted widths of the $\piz$ peaks in data and
simulation. The resolution is varied in the unfolding by $\pm10\%$,
resulting in a systematic uncertainty of less than $1\%$ in every
$\pt$ interval. The efficiency correction uncertainty arises from the
finite size of the simulated data samples and results in a global
uncertainty of about $1\%\mbox{--}2\%$. The luminosity has been measured
in $pp$ collisions with a precision of $2\%$ and in $p$-Pb
collisions with a precision of $2.6\%$ in the forward configuration
and $2.5\%$ in the backward configuration. The luminosity uncertainty
is $50\%$ correlated between datasets. The differential cross sections
have an additional $4\%$ uncertainty due to uncertainties in the
detector material budget. This uncertainty is fully correlated between
datasets and cancels in the nuclear modification factor.

{\setlength{\tabcolsep}{2pt}

\begin{table}[tb]
 \centering
 
 \caption{Relative systematic and statistical uncertainties in $d\sigma/d\pt$
 and $R_{p\mbox{-}{\rm Pb}}$ in percent. The ranges correspond to the minimum
 and maximum values of the associated uncertainties across all $\pt$ intervals
 and both $\eta_{\rm c.m.}$ regions. The $d\sigma/d\pt$ ranges cover
 the uncertainties for each of the $p$-Pb and $pp$ samples. All sources of
 systematic uncertainty are fully correlated across $\pt$ intervals.}

 \begin{tabular}{l c rcl c rcl}
   \hline
   Source && \multicolumn{3}{c}{$d\sigma/d\pt\,[\%]$} && \multicolumn{3}{c}{$R_{p\mbox{-}{\rm Pb}}\,[\%]$}\\
   \hline
   Fit model &~~~& 2.0 & -- & 12.6 &~~~& 0.9 & -- & 15.8 \\
   Unfolding && 0.3 & -- & 6.4 && 0.4 & -- & 6.4 \\
   Interpolation && & $-$ & && 0.9 & -- & 4.5 \\
   Material && & 4.0 & && & $-$ &  \\
   Efficiency && 1.3 & -- & 1.9 && 1.9 &  -- & 2.1 \\
   Luminosity && 2.0 & -- & 2.6 && 2.2 &  -- & 2.3 \\
   \hline
   Total systematic && 5.4 & -- & 15.0 && 4.3 & -- & 17.4\\
   \hline
   Statistical && 1.0 & -- & 9.6 && 1.4 & -- & 9.1\\
   \hline
 \end{tabular}
 \label{tab:systematics}
\end{table}

}

\begin{figure}[tb]
\centering
\includegraphics[width=0.55\textwidth]{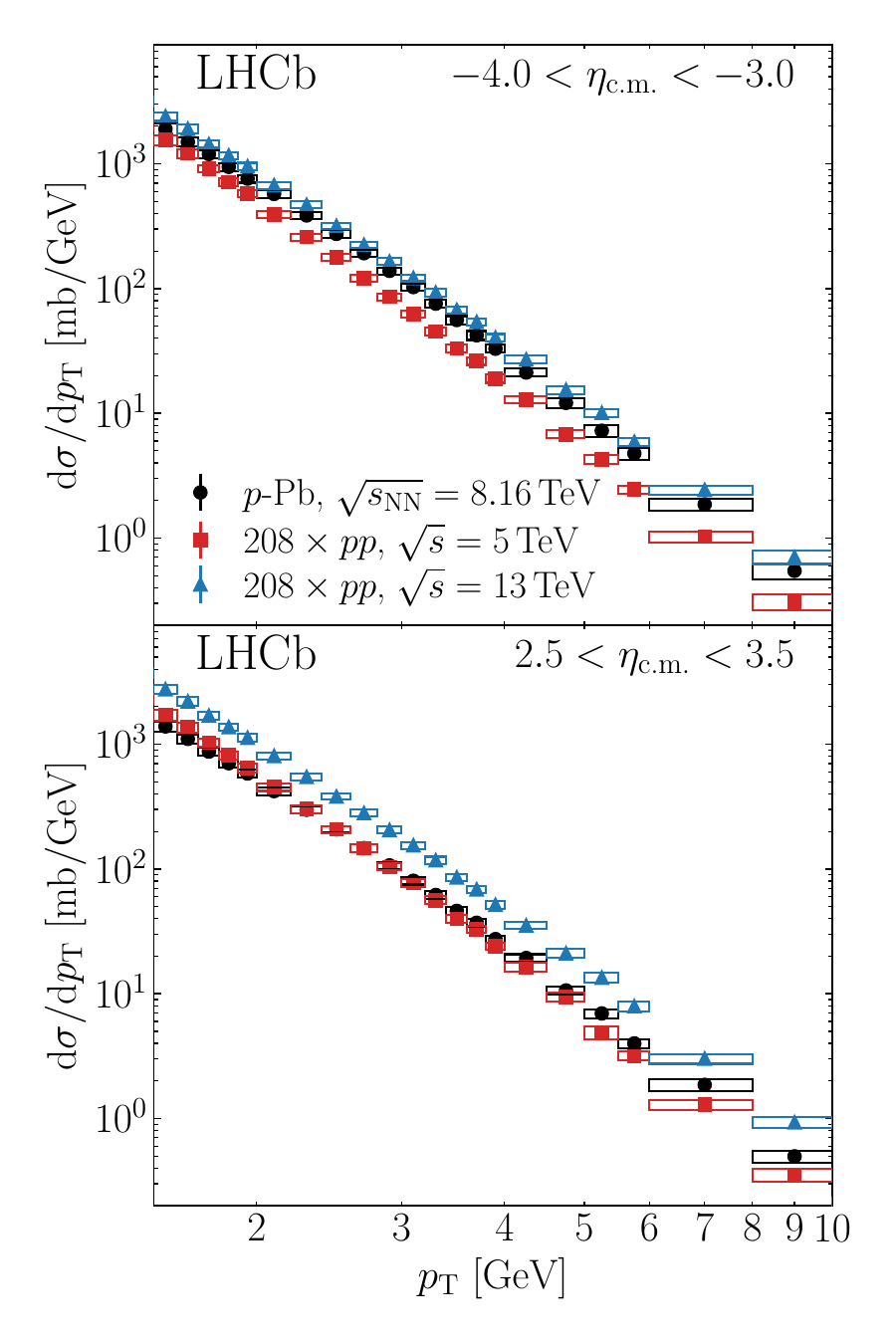}
\caption{
Measured $\piz$ differential cross sections versus $\pt$ in the (top) backward
and (bottom) forward $\eta_{\rm c.m.}$ regions.
Statistical uncertainties are shown by error bars, while systematic uncertainties are shown by boxes. The $pp$ cross sections are scaled by the atomic mass of the lead ion,
\mbox{$A=208$}.}
\label{fig:xsec}
\end{figure}

\begin{figure}[tb]
\centering
\includegraphics[width=\textwidth]{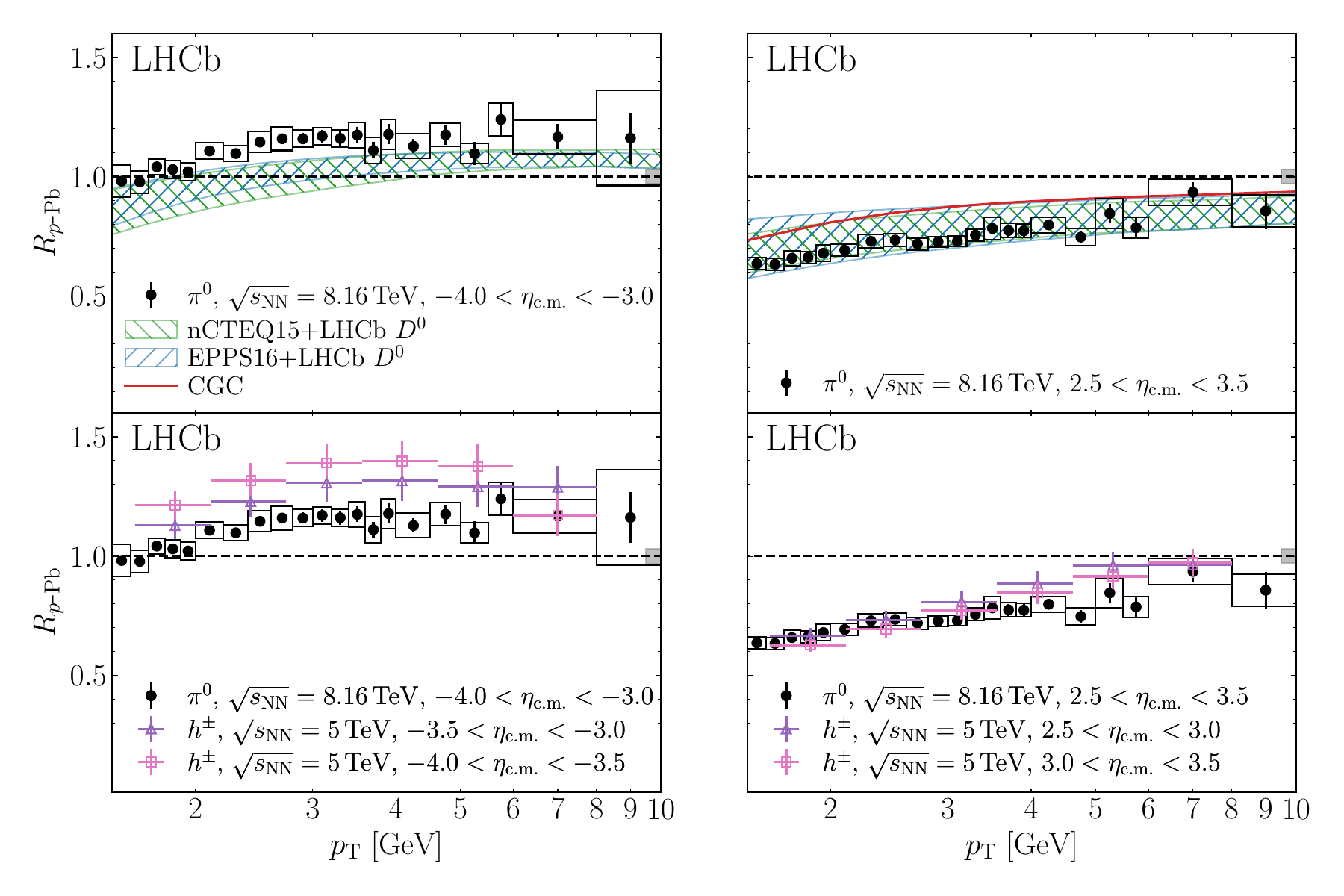}
\caption{
Measured $\piz$ nuclear modification factor in the (left)
backward and (right) forward $\eta_{\rm c.m.}$ regions. Error bars show
the statistical uncertainty, while the open boxes show the \pt-dependent systematic
uncertainties. The solid gray boxes show the
overall normalization uncertainties from the luminosity estimate and
efficiency correction factors. The results are compared to (top)
theoretical predictions~\cite{Helenius:2014qla,Eskola:2019bgf,Lappi:2013zma} and
(bottom) to charged-particle data from Ref.~\cite{LHCb-PAPER-2021-015}. The hatched regions show the NPDF uncertainties of the PQCD calculations. The vertical error bars on the
charged-particle results show the combined systematic and statistical
uncertainties.}
\label{fig:rpa}
\end{figure}

The fully corrected $\piz$ differential cross sections and nuclear modification
factor are shown in Figs.\,\ref{fig:xsec} and \ref{fig:rpa}, respectively. The
nuclear modification factor shows a Cronin-like enhancement at backward
pseudorapidity and a strong suppression at forward pseudorapidity. These
measurements are compared to next-to-leading order PQCD
calculations~\cite{Helenius:2014qla} using the EPPS16~\cite{Eskola:2016oht} and
nCTEQ15~\cite{Kovarik:2015cma} NPDF sets and the DSS14 $\piz$ fragmentation
functions~\cite{deFlorian:2014xna}. The NPDFs have been reweighted to
incorporate LHCb $D^0$ production data
\cite{Eskola:2019bgf,LHCb-PAPER-2017-015,Supp}, resulting in considerably
smaller uncertainties than calculations using the default EPPS16 and nCTEQ15
NPDFs (see Supplemental Material~\cite{Supp}). The measurement uncertainties in
the forward region are much smaller than the NPDF uncertainties, indicating that
this measurement can provide powerful constraints on NPDFs at low $x$. In
addition, the forward results present tension with the CGC
calculation~\cite{Lappi:2013zma}. The enhancement in the backward direction
between $2$ and $4~\gev$ is larger than predicted by the PQCD calculation,
suggesting that effects not described by NPDFs contribute to the enhancement.

The $\piz$ nuclear modification factor is also compared to the charged-particle
nuclear modification factor measured by the LHCb experiment in $p$-Pb collisions
at \mbox{$\sqrt{s_{\rm NN}}=5\,\tev$}~\cite{LHCb-PAPER-2021-015}. The forward
$\piz$ measurement agrees with the charged-particle measurement, and the
enhancement at backward $\eta_{\rm c.m.}$ is smaller than that seen for charged
particles at the LHCb experiment. Because the charged-particle measurement
includes heavier mesons and baryons, this ordering could indicate a
mass-dependent enhancement consistent with radial flow or baryon enhancement
from final-state recombination~\cite{Hwa:2004zd}. Studies of protons and heavier
unflavored mesons, such as $\eta$ and $\eta'$ mesons, could help differentiate
between these explanations.

In conclusion, the $\piz$ nuclear modification factor is measured at
forward and backward rapidities in $p$-Pb collisions at
\mbox{$\sqrt{s_{\rm NN}}=8.16\,\tev$}. The measured nuclear modification
factor has a total uncertainty of less than $6\%$ in most $\pt$
intervals, which will provide strong constraints on models of nuclear
structure and particle production in heavy ion collisions. In
particular, the forward measurement is sensitive to NPDFs for $x$ as
low as $10^{-6}$ and can provide useful constraints in future NPDF
analyses. Furthermore, the backward measurement shows the first
evidence of enhanced $\piz$ production in proton-ion collisions at the
LHC. These measurements will help constrain nuclear parton densities
and models of collectivity in small-collision systems, as well as
explain the origin of Cronin enhancement at the LHC.

\section*{Acknowledgements}

\noindent 
We thank Kari Eskola, Ilkka Helenius, Heikki M\"antysaari, Petja Paakkinen, and Hannu Paukkunen for providing theoretical predictions.
We express our gratitude to our colleagues in the CERN
accelerator departments for the excellent performance of the LHC. We
thank the technical and administrative staff at the LHCb
institutes.
We acknowledge support from CERN and from the national agencies:
CAPES, CNPq, FAPERJ and FINEP (Brazil); 
MOST and NSFC (China); 
CNRS/IN2P3 (France); 
BMBF, DFG and MPG (Germany); 
INFN (Italy); 
NWO (Netherlands); 
MNiSW and NCN (Poland); 
MEN/IFA (Romania); 
MICINN (Spain); 
SNSF and SER (Switzerland); 
NASU (Ukraine); 
STFC (United Kingdom); 
DOE NP and NSF (USA).
We acknowledge the computing resources that are provided by CERN, IN2P3
(France), KIT and DESY (Germany), INFN (Italy), SURF (Netherlands),
PIC (Spain), GridPP (United Kingdom), 
CSCS (Switzerland), IFIN-HH (Romania), CBPF (Brazil),
Polish WLCG  (Poland) and NERSC (USA).
We are indebted to the communities behind the multiple open-source
software packages on which we depend.
Individual groups or members have received support from
ARC and ARDC (Australia);
Minciencias (Colombia);
AvH Foundation (Germany);
EPLANET, Marie Sk\l{}odowska-Curie Actions and ERC (European Union);
A*MIDEX, ANR, IPhU and Labex P2IO, and R\'{e}gion Auvergne-Rh\^{o}ne-Alpes (France);
Key Research Program of Frontier Sciences of CAS, CAS PIFI, CAS CCEPP, 
Fundamental Research Funds for the Central Universities, 
and Sci. \& Tech. Program of Guangzhou (China);
GVA, XuntaGal, GENCAT and Prog.~Atracci\'on Talento, CM (Spain);
SRC (Sweden);
the Leverhulme Trust, the Royal Society
 and UKRI (United Kingdom).

\addcontentsline{toc}{section}{References}

\ifx\mcitethebibliography\mciteundefinedmacro
\PackageError{LHCb.bst}{mciteplus.sty has not been loaded}
{This bibstyle requires the use of the mciteplus package.}\fi
\providecommand{\href}[2]{#2}

\newpage
\clearpage
\newpage
\setcounter{equation}{0}
\setcounter{figure}{0}
\setcounter{table}{0}
\setcounter{section}{0}
\setcounter{page}{1}
\makeatletter
\renewcommand{\theequation}{S\arabic{equation}}
\renewcommand{\thefigure}{S\arabic{figure}}
\renewcommand{\thetable}{S\arabic{table}}
\renewcommand{\thepage}{S\arabic{page}}
\newcommand\ptwiddle[1]{\mathord{\mathop{#1}\limits^{\scriptscriptstyle(\sim)}}}

\begin{center}
\textbf{\Large Supplemental Material for LHCb-PAPER-2021-053} \\
\vspace{0.15in}
\textbf{\large \it Nuclear modification factor of neutral pions in the forward and backward regions in $p{\rm Pb}$ collisions} \\
\end{center}

\subsection*{Additional Comparisons to Theory}

\begin{figure}[h!]
  \centering
  \includegraphics[width=\textwidth]{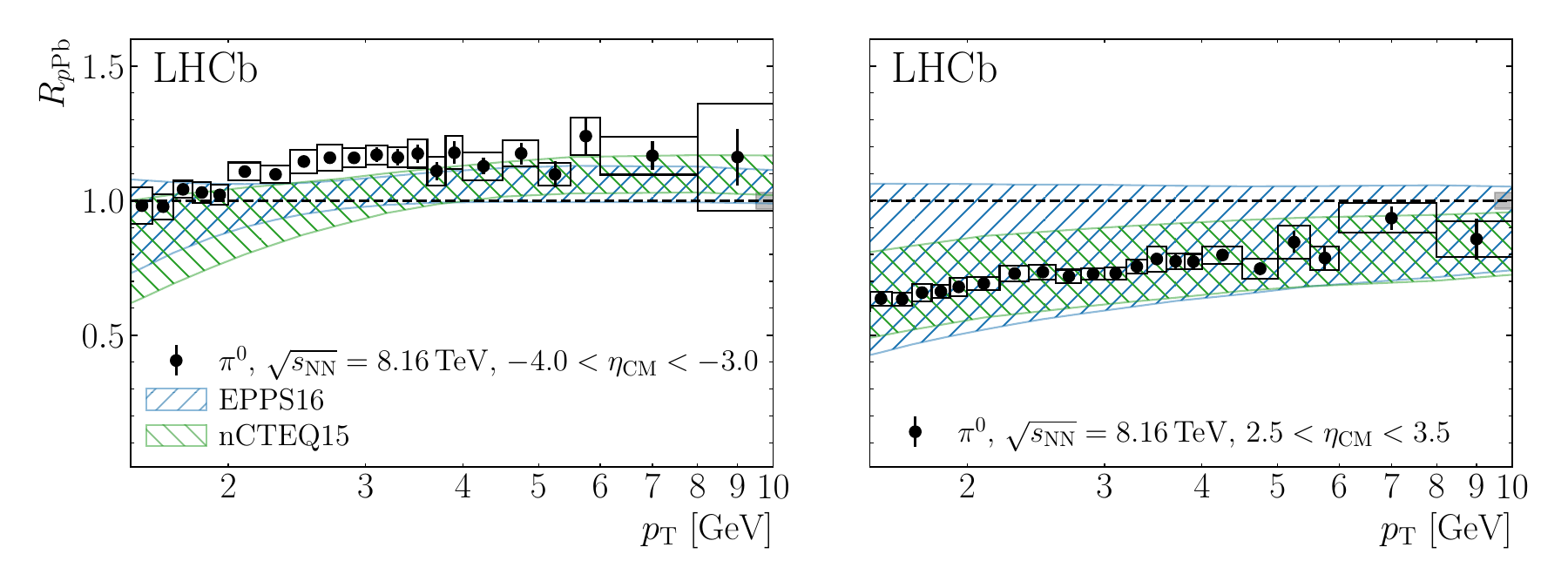}

  \caption{Measured $\pi^0$ nuclear modification factor in the
  (left) backward and (right) forward $\eta_{\rm CM}$ regions. The
  error bars show the statistical uncertainties, while the open boxes
  show the \pt-dependent systematic uncertainties. The solid
  black boxes show the overall normalization uncertainties from the
  luminosity estimate and efficiency correction factors. The
  measurements are compared to NLO PQCD
  predictions~\cite{Helenius:2014qla} calculated using the EPPS16~\cite{Eskola:2016oht} and
  nCTEQ15~\cite{Kovarik:2015cma} NPDF sets.}

  \label{fig:supp:rw}
\end{figure}

Fig.~\ref{fig:supp:rw}
shows comparisons of the measured $\piz$ nuclear modification factor
to NLO PQCD predictions calculated using the nominal EPPS16~\cite{Eskola:2016oht} and nCTEQ15~\cite{Kovarik:2015cma} NPDF sets. These NPDF analyses do not incorporate LHCb $D^0$ data~\cite{LHCb-PAPER-2017-015}, which provide strong constraints at low $x$. As a result, calculations using these NPDF sets have much larger uncertainties than those incorporating LHCb measurements.

\newpage
\centerline
{\large\bf LHCb collaboration}
\begin
{flushleft}
\small
R.~Aaij$^{32}$\lhcborcid{0000-0003-0533-1952},
A.S.W.~Abdelmotteleb$^{50}$\lhcborcid{0000-0001-7905-0542},
C.~Abellan~Beteta$^{44}$,
F.~Abudin{\'e}n$^{50}$\lhcborcid{0000-0002-6737-3528},
T.~Ackernley$^{54}$\lhcborcid{0000-0002-5951-3498},
B.~Adeva$^{40}$\lhcborcid{0000-0001-9756-3712},
M.~Adinolfi$^{48}$\lhcborcid{0000-0002-1326-1264},
H.~Afsharnia$^{9}$,
C.~Agapopoulou$^{13}$\lhcborcid{0000-0002-2368-0147},
C.A.~Aidala$^{77}$\lhcborcid{0000-0001-9540-4988}
% ...

{\footnotesize \it

% ...
$^{9}$Universit{\'e} Clermont Auvergne, CNRS/IN2P3, LPC, Clermont-Ferrand, France\\
% ...
$^{13}$LPNHE, Sorbonne Universit{\'e}, Paris Diderot Sorbonne Paris Cit{\'e}, CNRS/IN2P3, Paris, France\\
% ...
$^{32}$Nikhef National Institute for Subatomic Physics, Amsterdam, Netherlands\\
% ...
$^{40}$Instituto Galego de F{\'\i}sica de Altas Enerx{\'\i}as (IGFAE), Universidade de Santiago de Compostela, Santiago de Compostela, Spain\\
% ...
$^{44}$Physik-Institut, Universit{\"a}t Z{\"u}rich, Z{\"u}rich, Switzerland\\
% ...
$^{48}$H.H. Wills Physics Laboratory, University of Bristol, Bristol, United Kingdom\\
% ...
$^{50}$Department of Physics, University of Warwick, Coventry, United Kingdom\\
% ...
$^{54}$Oliver Lodge Laboratory, University of Liverpool, Liverpool, United Kingdom\\
% ...
$^{62}$Syracuse University, Syracuse, NY, United States\\
% ...
$^{77}$University of Michigan, Ann Arbor, MI, United States, associated to $^{62}$\\
\bigskip
% ...
}
\end{flushleft}

\end{document}